\documentclass[conference,10pt]{IEEEtran}

\usepackage[utf8]{inputenc}
\usepackage[T1]{fontenc}
\usepackage{textcomp}

\usepackage{algorithm}
\usepackage{algpseudocode}

\usepackage{amsmath,amssymb,amsfonts}

\usepackage[table]{xcolor} 

\usepackage{array}
\usepackage{booktabs}
\usepackage{multirow}
\usepackage{tabularx}

\usepackage{graphicx}
\usepackage{tikz}
\usetikzlibrary{shapes.geometric, arrows.meta, positioning, fit, calc,
    backgrounds, decorations.markings, patterns
}
\usepackage{pgfplots}
\pgfplotsset{compat=1.18}

\usepackage{enumitem}
\usepackage{listings}
\usepackage{comment}

\usepackage{float}
\usepackage{balance}
\usepackage{enumitem}

\usepackage{cite}

\usepackage{hyperref}
\usepackage{cleveref}

\definecolor{colorgray}{gray}{0.92}
\definecolor{darkblue}{RGB}{0,51,102}
\definecolor{midblue}{RGB}{70,130,180}
\definecolor{darkred}{RGB}{160,40,40}
\definecolor{darkgreen}{RGB}{0,100,60}
\definecolor{rowgray}{RGB}{240,240,245}
\definecolor{supportgreen}{RGB}{0,105,60}
\definecolor{supportamber}{RGB}{160,100,0}
\definecolor{supportgray}{RGB}{100,100,100}

\newcommand{\FullSupport}{%
  \textcolor{supportgreen}{$\checkmark$}}
\newcommand{\PartialSupport}{%
  \textcolor{supportamber}{$\circ$}}
\newcommand{\NotPrimary}{%
  \textcolor{supportgray}{--}}

\lstdefinelanguage{dapl}{
    basicstyle=\ttfamily\scriptsize,
    keywordstyle=\color{darkblue}\bfseries,
    commentstyle=\color{darkgreen!70!black}\itshape,
    stringstyle=\color{darkred!80!black},
    keywords={actor_id, subject_token, tenant_id, roles, enforcement_mode, policy_ids, correlation_id, statement_id, cursor_id, read_only, max_rows, max_bytes, timeout_ms, schema, payload, lineage, resources, rows, bytes, elapsed_ms, true, false, null},
    morestring=[b]",
    morecomment=[l]{//},
    showstringspaces=false,
    breaklines=true,
    frame=single,
    framerule=0.3pt,
    rulecolor=\color{darkblue!40},
    backgroundcolor=\color{darkblue!3},
    tabsize=2,
    columns=fullflexible
}

\title{{\spaceskip=0.25em\relax
SCOPE-DT: Dependency-Aware Authority Containment and Recovery for Agentic Digital Twins}}

\author{
\IEEEauthorblockN{\makebox[0.30\textwidth][c]{1\textsuperscript{st} Joyjit Roy}}
\IEEEauthorblockA{
\makebox[0.30\textwidth][c]{\textit{IEEE Senior Member}} \\
\makebox[0.30\textwidth][c]{Independent Researcher} \\
\makebox[0.30\textwidth][c]{Austin, Texas, USA} \\
\makebox[0.30\textwidth][c]{joyjit.roy.tech@gmail.com}}
}

\begin{document}

\maketitle

\begin{abstract}
Agentic digital twins require execution authority that adapts to dependency failures without unnecessarily suspending independent operations. This paper presents SCOPE-DT, a framework that integrates dependency-aware containment, epoch-bound execution permits, and incident-specific recovery obligations. Action requirements address mandatory, alternative, and quorum dependencies, while recovery processes separate evidence renewal, compensation, and reconciliation from normal execution. Across 288 evaluation runs spanning three domains and eight incident scenarios, SCOPE-DT recorded mean unsafe-execution rates of 0.66--0.94\% per proposal and completed 96.35--96.69\% of unaffected safe proposals. The mean premature-restoration rate was 0.93\%, compared with 5.22\% for full policy reevaluation. Permit tests and sensitivity analysis revealed residual enforcement failures and reliance on accurate dependency records. Cost-function estimates indicate lower reassessment costs than full policy reevaluation for larger graph sizes. These results support selective containment and obligation-guided recovery while highlighting remaining deployment challenges.
\end{abstract}

\vspace{-6pt}
\section{Introduction}
\label{sec:intro}

\textbf{Agentic Digital Twins.}
Digital twins represent operational state and transitions to support monitoring and decision-making~\cite{voas2025digitaltwin}. Autonomous agents extend these capabilities to operational actions. AgentSOC demonstrates how agentic reasoning supports security response planning through feasibility validation and policy constraints~\cite{roy2026agentsoc}. However, execution authority depends on the reliability of supporting state, evidence, and services. Dependency failures require determining which actions must be restricted and which can safely continue.

\textbf{Dependency Failures and Containment.}
Dependency failures may disrupt several actions but leave independent operations viable. Decentralized multi-agent swarms enable local threat validation and quarantine in industrial IoT~\cite{singh2026dmas}. Microrebooting limits disruption similarly by recovering individual application components~\cite{candea2004microreboot}. In agentic digital twins, global suspension can unnecessarily interrupt independent work, whereas fixed local isolation may overlook shared or transitive dependencies. Effective containment requires identifying which actions can no longer satisfy their execution requirements following a dependency failure.

\textbf{Execution Authority and Recovery.}
Runtime assurance limits autonomous control when monitored safety conditions are violated~\cite{mehmood2022blackbox}. Cedar supports policy-based authorization~\cite{cutler2024cedar}, and digital-twin resilience frameworks analyze failure propagation and support adaptation~\cite{parri2021resilient}. For agentic digital twins, these capabilities must be connected to action-specific dependencies and active execution permits. Restoration must identify recovered dependencies, evidence that requires renewal, and unresolved workflow outcomes that need reconciliation before execution authority can safely resume.

\textbf{Proposed Approach.}
This paper introduces SCOPE-DT, a dependency-aware framework for authority containment and recovery in agentic digital twins. The main contributions are as follows:

\begin{itemize}[leftmargin=*, topsep=0pt, itemsep=0pt, parsep=0pt]
\item A dependency model that evaluates action-specific AND, OR, and quorum requirements to identify which actions lose execution authority after an incident, while preserving independently supported operations.
\item Execution permits bound to the epochs of their supporting dependencies, so that only relevant dependency changes invalidate a permit.
\item Incident-specific recovery obligations that gate the restoration of normal execution authority, accounting for renewed evidence and unresolved workflow effects.
\item An evaluation across three domains and eight incident scenarios, comparing containment, permit enforcement, and recovery against four baselines.
\end{itemize}
\section{Background and Related Work}
\label{sec:related_work}

\subsection{Digital Twins and Operational Dependencies}

\textbf{NIST IR 8356} report discusses security and trust challenges related to digital-twin representations and their operational uses~\cite{voas2025digitaltwin}. \textbf{JARVIS} links digital twins to executable failure models to support failure propagation analysis and system adaptation~\cite{parri2021resilient}. \textbf{CPSRA} uses state dependency graphs to analyze cascading attack paths and identify mitigation points in cyber-physical systems~\cite{adamos2024resilience}. These approaches lay the foundation for dependency analysis and adaptation.

\subsection{Runtime Assurance and Execution Authorization}

\textbf{Black-Box Simplex} switches control between advanced and baseline controllers through runtime safety checks~\cite{mehmood2022blackbox}. \textbf{Cedar} separates authorization policies from application logic and supports role-, attribute-, and relationship-based access decisions~\cite{cutler2024cedar}. \textbf{Zanzibar} maintains causal ordering between authorization decisions and updates to policies or object contents~\cite{pang2019zanzibar}. \textbf{AgentSOC} integrates contextual reasoning, feasibility validation, and policy-constrained response planning for security operations~\cite{roy2026agentsoc}. \textbf{CaMeL} separates model reasoning from tool-call policy enforcement to constrain agent actions~\cite{debenedetti2026camel}. These approaches support runtime enforcement, policy-driven decisions, and authorization consistency.

\subsection{Fault Containment and Workflow Recovery}

\textbf{DMAS} uses decentralized agents in industrial IoT to coordinate local threat validation and quarantine~\cite{singh2026dmas}. \textbf{Microreboot} restores faulty application components while preserving unaffected operations~\cite{candea2004microreboot}. \textbf{Sagas} organize long-running workflows into committed steps with compensating transactions for partial execution~\cite{garciamolina1987sagas}. These mechanisms enable localized response and recovery. SCOPE-DT links containment to action-specific dependencies and separates normal execution authority from permissions for compensation and reconciliation. Restoration requires fulfilling incident-specific obligations before affected actions resume. Table~\ref{tab:related_work_comparison} summarizes these approaches.

\begin{table}[t]
\caption{Comparison of Related Approaches and SCOPE-DT}
\label{tab:related_work_comparison}
\centering
\footnotesize
\setlength{\tabcolsep}{2pt}
\renewcommand{\arraystretch}{1.0}

\begin{tabular}{
@{}
p{0.36\columnwidth}                         
>{\centering\arraybackslash}p{0.09\columnwidth} 
>{\centering\arraybackslash}p{0.09\columnwidth} 
>{\centering\arraybackslash}p{0.09\columnwidth} 
>{\centering\arraybackslash}p{0.09\columnwidth} 
>{\centering\arraybackslash}p{0.09\columnwidth} 
>{\centering\arraybackslash}p{0.09\columnwidth} 
@{}
}
\hline
\rowcolor{colorgray}
\textbf{Approach} &
\textbf{DT} &
\textbf{Dep.} &
\textbf{Auth.} &
\textbf{Cont.} &
\textbf{WF} &
\textbf{Rec.} \\
\hline

\rowcolor{white}
JARVIS~\cite{parri2021resilient}
& \FullSupport & \FullSupport & \NotPrimary
& \PartialSupport & \NotPrimary & \FullSupport \\

\rowcolor{colorgray}
CPSRA~\cite{adamos2024resilience}
& \PartialSupport & \FullSupport & \NotPrimary
& \PartialSupport & \NotPrimary & \NotPrimary \\

\rowcolor{white}
Black-Box Simplex~\cite{mehmood2022blackbox}
& \NotPrimary & \PartialSupport & \FullSupport
& \PartialSupport & \NotPrimary & \PartialSupport \\

\rowcolor{colorgray}
Cedar~\cite{cutler2024cedar}
& \NotPrimary & \PartialSupport & \FullSupport
& \PartialSupport & \NotPrimary & \NotPrimary \\

\rowcolor{white}
Zanzibar~\cite{pang2019zanzibar}
& \NotPrimary & \PartialSupport & \FullSupport
& \PartialSupport & \NotPrimary & \NotPrimary \\

\rowcolor{colorgray}
AgentSOC~\cite{roy2026agentsoc}
& \NotPrimary & \PartialSupport & \PartialSupport
& \PartialSupport & \NotPrimary & \NotPrimary \\

\rowcolor{white}
DMAS~\cite{singh2026dmas}
& \NotPrimary & \NotPrimary & \NotPrimary
& \FullSupport & \NotPrimary & \NotPrimary \\

\rowcolor{colorgray}
Microreboot~\cite{candea2004microreboot}
& \NotPrimary & \PartialSupport & \NotPrimary
& \FullSupport & \PartialSupport & \FullSupport \\

\rowcolor{white}
Sagas~\cite{garciamolina1987sagas}
& \NotPrimary & \PartialSupport & \NotPrimary
& \PartialSupport & \FullSupport & \FullSupport \\

\rowcolor{colorgray}
\textbf{SCOPE-DT}
& \FullSupport & \FullSupport & \FullSupport
& \FullSupport & \FullSupport & \FullSupport \\
\hline
\end{tabular}

\vspace{1mm}
\parbox{\columnwidth}{
\scriptsize
\textit{DT}: operational digital twin;
\textit{Dep.}: dependency modeling;
\textit{Auth.}: runtime authorization;
\textit{Cont.}: selective containment;
\textit{WF}: partial-workflow handling;
\textit{Rec.}: operational recovery.
\FullSupport{} = explicit;
\PartialSupport{} = partial or indirect;
\NotPrimary{} = not a primary focus.
}
\vspace{-3mm}
\end{table}
\section{System Model and Problem Formulation}
\label{sec:system_model}

\subsection{Actions and Operational Dependencies}
\label{sec:action_dependencies}

An agentic digital twin provides an operational model of a physical or digital system. Autonomous agents propose actions, and an independent authorization service evaluates their execution requirements before an execution service performs permitted actions. Let $\mathcal{A}$ denote the set of actions. Each action $a \in \mathcal{A}$ has a target scope and a dependency set $\mathcal{D}_a$ comprising operational state, evidence, services, infrastructure, and workflow conditions relevant to execution. Dependencies may be shared among actions or rely on other dependencies, so a single failure can affect multiple execution paths.

Let $v_d(t) \in \{0,1\}$ indicate whether dependency $d$ satisfies its validity conditions at time $t$. Following Boolean gate semantics from fault-tree analysis~\cite{ruijters2015faulttree}, a requirement over dependency group $Q$ is evaluated as

\begin{equation}
g_{Q,k}(t)
= \mathbf{1}\!\left[\sum_{d \in Q} v_d(t) \geq k\right],
\quad 1 \leq k \leq |Q|,
\label{eq:dependency_gate}
\end{equation}

where $\mathbf{1}[\cdot]$ equals one when its condition holds and zero otherwise. AND requires $k=|Q|$, OR requires $k=1$, and a quorum uses its specified threshold $k$. These gates may be nested to form the action-specific predicate $\phi_a(t)$. A failure removes dependency support only when it causes $\phi_a(t)=0$. Valid alternatives or a sufficient remaining quorum can preserve support. Policy and permit checks remain necessary.

\subsection{Containment and Recovery Requirements}
\label{sec:containment_requirements}

\begin{table}[t]
\caption{System Model Notation}
\label{tab:notation}
\centering
\footnotesize
\setlength{\tabcolsep}{3pt}
\renewcommand{\arraystretch}{1.0}

\begin{tabular}{@{}ll@{}}
\hline
\rowcolor{colorgray}
\textbf{Symbol} & \textbf{Definition} \\
\hline

\rowcolor{white}
$\mathcal{A},\,a$ & Action set and a member action \\

\rowcolor{colorgray}
$\mathcal{D}_a$ & Dependencies relevant to action $a$ \\

\rowcolor{white}
$v_d(t)$ & Binary validity of dependency $d$ at time $t$ \\

\rowcolor{colorgray}
$Q,\,k$ & Dependency group and required threshold \\

\rowcolor{white}
$g_{Q,k}(t)$ & Threshold gate for group $Q$ at time $t$ \\

\rowcolor{colorgray}
$\phi_a(t)$ & Dependency predicate for action $a$ at time $t$ \\

\rowcolor{white}
$I$ & Detected incident affecting dependency validity \\

\rowcolor{colorgray}
$\mathcal{I}_a(t)$ & Incidents with obligations applicable to action $a$ \\

\rowcolor{white}
$\mathcal{O}_{I,a}$ & Recovery obligations for action $a$ from incident $I$ \\

\rowcolor{colorgray}
$\nu_o(t)$ & Verified fulfillment of obligation $o$ \\

\rowcolor{white}
$p_a(t)$ & Current policy compliance \\

\rowcolor{colorgray}
$b_a(t)$ & Current permit validity \\

\rowcolor{white}
$\chi_a(t)$ & Combined check for normal execution \\

\rowcolor{colorgray}
$\mathcal{S}_a$ & Dependency support selected at permit issuance \\

\rowcolor{white}
$\pi_a$ & Execution permit for action $a$ \\

\rowcolor{colorgray}
$e_d(t)$ & Current epoch of dependency $d$ \\

\rowcolor{white}
$\widehat{e}_{a,d}$ & Epoch of $d$ recorded at permit issuance \\

\hline
\end{tabular}
\vspace{-3mm}
\end{table}

A detected incident $I$ identifies dependencies that no longer satisfy their validity conditions. Containment requires reevaluating the requirements of all actions that depend, directly or indirectly, on these dependencies. Actions for which $\phi_a(t)=0$ must lose normal execution authority. Actions whose predicates remain satisfied stay eligible, subject to policy and permit checks. Eligibility alone does not keep an existing permit valid. Its dependency bindings must remain valid before any effectful step proceeds.

Restoring a dependency alone does not resolve interrupted workflow effects. Distributed workflows must track uncertain outcomes and outstanding commitments~\cite{helland2007life}. For incident $I$ and action $a$, let $\mathcal{O}_{I,a}$ denote recovery obligations based on the incident cause and workflow state, such as evidence renewal, compensation, or reconciliation. Recovery operations require separately scoped permissions with their own execution requirements. Let $\mathcal{I}_a(t)$ contain incidents with obligations still applicable to $a$, and let $\nu_o(t)=1$ indicate verified fulfillment of obligation $o$ using currently valid evidence. With $p_a(t)$ and $b_a(t)$ denoting policy compliance and permit validity, including dependency bindings, normal execution requires

\vspace{-10pt}
\begin{equation}
\chi_a(t)=
\begin{cases}
1, &
\begin{aligned}
&\phi_a(t)=p_a(t)=b_a(t)=1,\\
&\nu_o(t)=1,\ 
  \forall I \in \mathcal{I}_a(t),\
  \forall o \in \mathcal{O}_{I,a},
\end{aligned}\\[2pt]
0, & \text{otherwise}.
\end{cases}
\label{eq:execution_eligibility}
\end{equation}

Thus, recovery from one incident cannot restore normal authority if obligations from another applicable incident remain unresolved. When no recovery obligations apply, the recovery condition is satisfied by default. Table~\ref{tab:notation} summarizes the notation.
\section{Framework and System Architecture}
\label{sec:arch}

\subsection{Architecture and Enforcement Boundaries}
\label{sec:architecture}

\begin{figure*}[!t]
    \centering
    \includegraphics[width=0.92\textwidth]{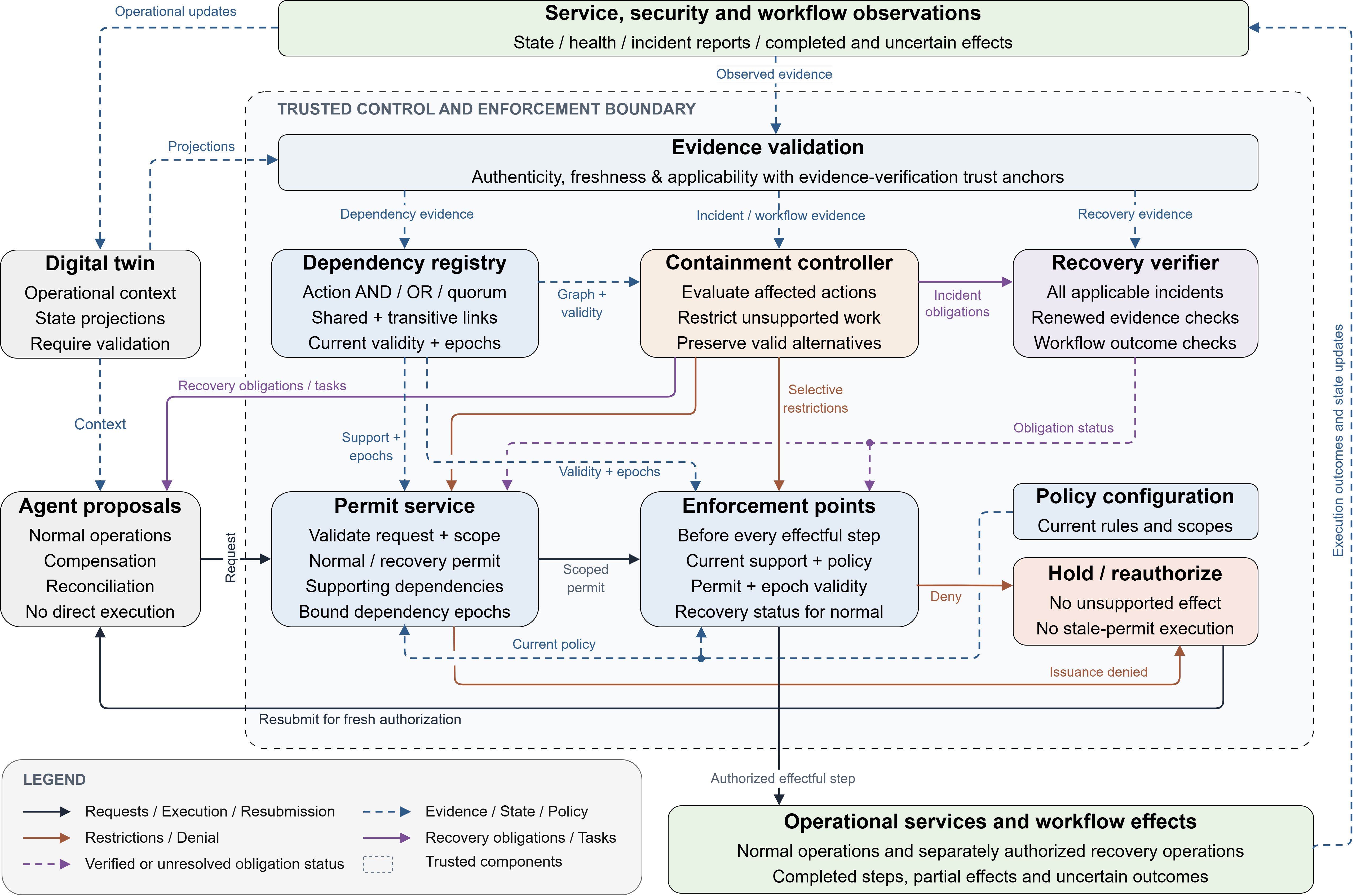}
    \caption{SCOPE-DT architecture showing evidence validation, dependency-aware containment, dependency-bound execution permits, and incident-specific recovery. Normal and recovery operations pass through shared enforcement points under separately scoped permissions.}
    \label{fig:scope_architecture}
    \vspace{-3mm}
\end{figure*}

SCOPE-DT separates action proposal, dependency assessment, permit issuance, and recovery verification, as shown in Fig.~\ref{fig:scope_architecture}. A dependency registry maintains action requirements, shared dependencies, and current validity and epoch records. A containment controller processes incident updates and identifies actions that require restriction or reauthorization. The permit service binds each execution permit to the action's supporting dependencies and their epochs. Enforcement points validate permits before each effectful step. A recovery verifier checks incident-specific obligations against current evidence before affected actions regain normal authority.

SCOPE-DT relies on the integrity of its control components, policy configuration, dependency registry, and evidence-verification mechanisms. Agent proposals and digital-twin projections require independent validation before supporting execution authority. Consistent with CaMeL's separation of model reasoning and tool-call policy enforcement~\cite{debenedetti2026camel}, every effectful request, including compensation and reconciliation, is routed through a non-bypassable enforcement point. Evidence is verified for authenticity, freshness, and relevance to the requested action. Execution is denied if required dependency support, permit validity, or recovery evidence cannot be established. These assumptions define the enforcement boundary for containment and restoration.

\subsection{Dependency-Aware Containment}
\label{sec:containment}

When an incident changes dependency validity, the registry updates the relevant records and advances their epochs. The containment controller follows reverse dependency links to identify actions whose requirements depend directly or indirectly on the changed dependencies. It reevaluates each affected action's predicate $\phi_a(t)$ using the current AND, OR, and quorum conditions and restricts actions whose predicates are no longer satisfied.

Containment is enforced at effectful-step boundaries. A permit invalidated after dispatch does not establish whether the step completed, failed, or produced partial effects. The controller therefore uses workflow observations to distinguish confirmed outcomes from unresolved effects and associates them with incident $I$. These records and the incident cause determine the recovery obligations $\mathcal{O}_{I,a}$ for each affected action. Compensation and reconciliation occur only with separately scoped recovery permissions. Algorithm~\ref{alg:containment} summarizes incident processing and selective containment.

\begin{algorithm}[t]
\footnotesize
\refstepcounter{algorithm}
\label{alg:containment}
\addcontentsline{loa}{algorithm}{\protect\numberline{\thealgorithm}Incident Processing and Selective Containment}
{\setlength{\fboxsep}{4pt}
\noindent\colorbox{black!12}{%
    \parbox{\dimexpr\linewidth-2\fboxsep\relax}{%
        \color{black}\textbf{Algorithm~\thealgorithm}\quad
        Incident Processing and Selective Containment
    }%
}\par}
\nointerlineskip
\hrule height 0.3pt
\smallskip

\begin{algorithmic}[1]
\Require Incident $I$, validated dependency updates, workflow observations
\Ensure Updated dependency records, obligations, and authority restrictions
\State $\Delta_I \gets$ dependencies changed by incident $I$
\State $\mathcal{A}_I \gets \Call{ReverseReachableActions}{\Delta_I}$
\State Stage updated validity and advanced epochs for $\Delta_I$
\For{each action $a \in \mathcal{A}_I$}
    \State Read confirmed and unresolved workflow effects for $a$
    \State Derive obligations from the incident cause and these effects
    \State Merge these obligations into $\mathcal{O}_{I,a}$
    \State Retain obligations from other applicable incidents
    \State Reevaluate $\phi_a(t)$ using the staged dependency state
    \If{$\phi_a(t)=0$}
        \State Restrict normal execution authority for $a$
    \Else
        \State Mark $a$ as dependency-supported
        \State Keep normal execution subject to $\chi_a(t)$
    \EndIf
    \State Invalidate permits whose bound dependency epochs changed
    \State Require fresh authorization before using replacement permits
\EndFor
\State Publish staged dependency \& recovery state as one revision
\State Route recovery requests through separately scoped authorization
\end{algorithmic}
\end{algorithm}

\subsection{Dependency-Bound Execution Permits}
\label{sec:execution_permits}

Each execution permit identifies the action, target scope, and authorized operation type, distinguishing normal execution from compensation and reconciliation. For action $a$, let $\mathcal{S}_a \subseteq \mathcal{D}_a$ denote the dependency support selected at issuance. This set satisfies $\phi_a(t)$ through the applicable mandatory, alternative, and quorum requirements, including transitive prerequisites. The permit records the current epoch for each dependency in $\mathcal{S}_a$. Changes outside this support set do not invalidate its epoch bindings. A change to a bound dependency's epoch requires reauthorization, even if another valid support set could satisfy the action's dependency predicate.

Before each effectful step, the enforcement point verifies the presented permit against the current control state. Let $\pi_a$ denote the permit for action $a$, $e_d(t)$ the current epoch of dependency $d$, and $\widehat{e}_{a,d}$ the epoch of $d$ recorded at issuance. Define $\operatorname{Valid}(\pi_a,a,t)$ to indicate that the permit is authentic, unexpired, not explicitly revoked, and matches the requested action, target scope, and operation type. Permit validity is then
\vspace{-5pt}
\begin{equation}
b_a(t)=
\begin{cases}
1, &
\begin{aligned}
&\operatorname{Valid}(\pi_a,a,t)=1,\\
&e_d(t)=\widehat{e}_{a,d},
\quad \forall d\in\mathcal{S}_a,
\end{aligned}\\[2pt]
0, & \text{otherwise}.
\end{cases}
\label{eq:permit_validity}
\end{equation}

A valid permit alone does not authorize execution. Each normal step must also satisfy the dependency, recovery, and policy conditions in $\chi_a(t)$. Compensation and reconciliation are assessed according to their specific execution requirements. The enforcement point must serialize final validation and step admission with relevant control-state updates, preventing an update from occurring between the check and admission. If the current state cannot be determined, the step is withheld.

\subsection{Incident-Specific Recovery}
\label{sec:recovery}

{\looseness=-1
SCOPE-DT restores normal execution authority through verified fulfillment of incident-specific recovery obligations. For each obligation $o \in \mathcal{O}_{I,a}$, the containment controller records the incident cause, affected dependency or workflow scope, required resolution, and evidence needed to verify completion. Evidence-related failures require renewed evidence from an acceptable source. Confirmed workflow effects may require compensation under the recovery policy, while uncertain outcomes require reconciliation before normal continuation. The recovery verifier evaluates completion evidence for authenticity, freshness, and relevance to the recorded obligation. A dependency repair fulfills an obligation only when the resulting evidence satisfies its recorded completion conditions.
\par}

Each obligation is verified separately across all applicable incidents. The verifier sets $\nu_o(t)=1$ only when current evidence satisfies the obligation's completion criteria. If the evidence becomes invalid, the verifier resets $\nu_o(t)$ to zero and requests renewed verification of the recorded outcome. Invalidated permits must be replaced with permits bound to current dependency epochs. Table~\ref{tab:recovery_obligations} summarizes typical incident causes, recovery obligations, and verification evidence.

\begin{table*}[t]
\caption{Incident-Specific Recovery Obligations and Evidence}
\label{tab:recovery_obligations}
\centering
\footnotesize
\setlength{\tabcolsep}{3pt}
\renewcommand{\arraystretch}{1.0}

\resizebox{\textwidth}{!}{%
\begin{tabular}{
@{}
>{\raggedright\arraybackslash}p{0.46\columnwidth}
>{\raggedright\arraybackslash}p{0.78\columnwidth}
>{\raggedright\arraybackslash}p{0.88\columnwidth}
@{}
}
\hline
\rowcolor{colorgray}
\textbf{Incident cause} &
\textbf{Recovery obligation} &
\textbf{Verification evidence} \\
\hline

\rowcolor{white}
Stale or missing evidence &
Renew the affected evidence &
Authentic, current evidence relevant to the action \\

\rowcolor{colorgray}
Compromised evidence source &
Establish an acceptable source and renew affected evidence &
Validated source trust and fresh evidence from that source \\

\rowcolor{white}
Dependency or quorum loss &
Restore sufficient valid support or validate an alternative &
Current dependency records demonstrating satisfied requirements \\

\rowcolor{colorgray}
Confirmed partial workflow effects &
Perform compensation required by recovery policy &
Compensation records and evidence of required postconditions \\

\rowcolor{white}
Uncertain workflow outcome &
Reconcile the outcome before normal continuation &
Authoritative status records correlated with the dispatched operation \\

\hline
\end{tabular}%
}
\vspace{-3mm}
\end{table*}
\section{Experimental Evaluation}
\label{sec:evaluation}

The evaluation covers containment, permit enforcement, recovery, mechanism sensitivity, and computational cost.

\subsection{Experimental Setup and Baselines}
\label{sec:experimental_setup}

The evaluation includes manufacturing-energy management, banking payments, and insurance claims across eight incident scenarios, totaling 288 runs. Each method receives identical proposal records and incident configurations per run. Table~\ref{tab:experimental_setup} summarizes the setup.

{\looseness=-1
The seeded generator assigns AND, OR, and quorum semantics using the probabilities in Table~\ref{tab:experimental_setup}. Each run selects a failed scope and incident severity, while scenario-specific exposure probabilities determine affected proposals. Ground-truth safety labels depend on affected status, sampled replacement support, and background invariant failures. Observed status includes errors determined by the configured graph-error rate. SCOPE-DT OR/quorum exception rules reference the generated safety labels and apply only to affected actions that are safe. They therefore do not alter unsafe-execution or unaffected-completion rates, but SCOPE-DT's safe-held rates represent lower bounds. The evaluation focuses on authorization-layer decisions. Agent behavior is outside its scope.

SCOPE-DT is compared with global suspension, fixed local isolation, directed reachability, and full policy reevaluation. Unsafe execution and unsafe escape are normalized by all proposals and unsafe proposals, respectively. Unnecessary holds are normalized by safe proposals, and unaffected completion by unaffected safe proposals. Results are mean per-run rates with run-level 95\% bootstrap confidence intervals. Permit, recovery, and scaling tests use separate generated cases (Sections~\ref{sec:permit_results}, \ref{sec:recovery_results}, and \ref{sec:scalability}).
\par}
\begin{table}[t]
\caption{Experimental Configuration and Incident Scenarios}
\label{tab:experimental_setup}
\centering
\footnotesize
\setlength{\tabcolsep}{3pt}
\renewcommand{\arraystretch}{1.08}

\resizebox{\columnwidth}{!}{%
\begin{tabular}{
@{}
>{\raggedright\arraybackslash}p{0.30\columnwidth}
>{\raggedright\arraybackslash}p{\dimexpr0.86\columnwidth-2\tabcolsep\relax}
@{}
}
\hline
\rowcolor{colorgray}
\textbf{Parameter} & \textbf{Configuration} \\
\hline

\rowcolor{white}
Domains & Manufacturing-energy, banking payments, and insurance claims \\

\rowcolor{colorgray}
Incident scenarios & Local staleness, redundant-source loss, quorum loss, shared-service outage, replica divergence, signer compromise, compound failure, and false-clear recovery \\

\rowcolor{white}
Evaluation runs & 12 per domain--scenario combination; 288 total \\

\rowcolor{colorgray}
Comparison methods & SCOPE-DT and four baselines; 1,440 method-level records \\

\rowcolor{white}
Action proposals & 74,719 total; 210--310 per run \\

\rowcolor{colorgray}
Graph-size settings & 180, 380, 740, 1,250, and 2,450 nodes \\

\rowcolor{white}
Scopes & 9--29 per run \\

\rowcolor{colorgray}
Generated dependency semantics & AND (49\%), OR (30\%), and quorum (21\%) \\

\rowcolor{white}
Paired comparisons & Shared proposal sets and incident configurations \\

\rowcolor{colorgray}
Confidence intervals & 95\%; 900 bootstrap resamples at the run level \\

\rowcolor{white}
Generator seed & 20261011 \\

\hline
\end{tabular}%
}
\vspace{-3mm}
\end{table}

\subsection{Containment Safety and Operational Continuity}
\label{sec:containment_results}

Table~\ref{tab:containment_results} compares safety and continuity across the three domains. SCOPE-DT recorded mean unsafe-execution rates of 0.66--0.94\% per proposal and completed 96.35--96.69\% of unaffected safe proposals. Global suspension blocked all proposals, eliminating unsafe execution while also stopping unaffected work. Fixed local isolation recorded unsafe-execution rates of 10.84--12.39\%, and directed reachability of 2.72--3.25\%. Full policy reevaluation achieved unsafe-execution rates similar to SCOPE-DT (0.67--0.97\%) while retaining higher unaffected completion (97.17--97.48\%).

\begin{figure*}[t]
\centering
\includegraphics[width=0.85\textwidth]{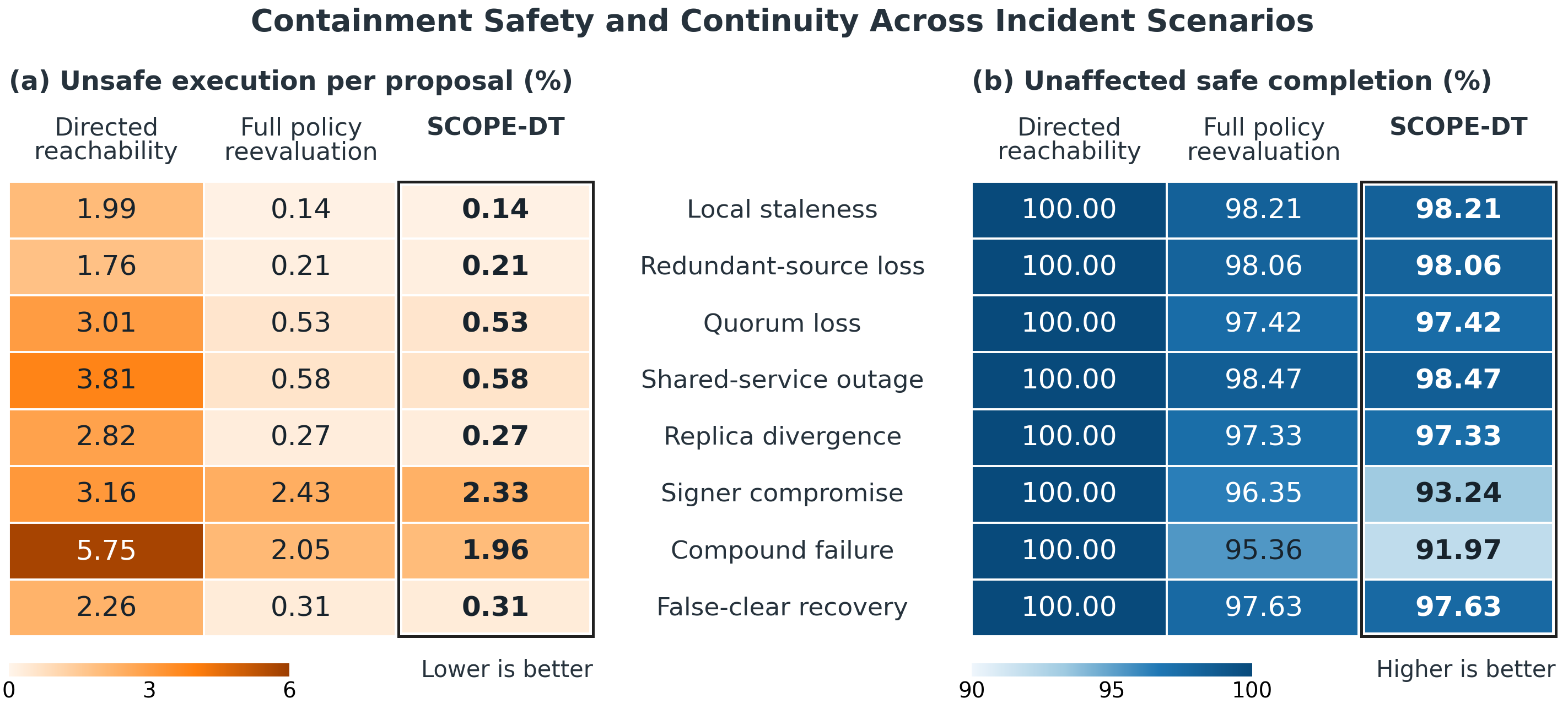}
\caption{Containment safety and operational continuity across eight incident scenarios. Cells show mean per-run percentages across 36 runs per scenario and method, covering three domains. Lower unsafe execution and higher unaffected completion are preferable.}
\label{fig:scenario_results}
\vspace{-4mm}
\end{figure*}

\begin{table}[t]
\caption{Containment Safety and Continuity Across Domains}
\label{tab:containment_results}
\centering
\footnotesize
\setlength{\tabcolsep}{2pt}
\renewcommand{\arraystretch}{1.2}

\newcommand{\scopevalue}[2]{%
    \shortstack{#1\\{\scriptsize[#2]}}%
}

\begin{tabularx}{\columnwidth}{
@{}
>{\centering\arraybackslash}p{30pt}
>{\raggedright\arraybackslash}p{48pt}
*{3}{>{\centering\arraybackslash}X}
@{}
}
\hline
\rowcolor{colorgray}
\textbf{Domain} &
\textbf{Method} &
\shortstack{\textbf{Unsafe /}\\\textbf{proposed}} &
\shortstack{\textbf{Safe}\\\textbf{held}} &
\shortstack{\textbf{Unaffected}\\\textbf{completion}} \\
\hline

\rowcolor{white}
\cellcolor{white}
& \shortstack[l]{Global\\suspension}
& \scopevalue{0.00}{0.00, 0.00}
& \scopevalue{100.00}{100.00, 100.00}
& \scopevalue{0.00}{0.00, 0.00} \\

\rowcolor{colorgray}
\cellcolor{white}
& \shortstack[l]{Fixed local\\isolation}
& \scopevalue{10.84}{9.61, 12.17}
& \scopevalue{5.14}{4.66, 5.67}
& \scopevalue{95.13}{94.65, 95.60} \\

\rowcolor{white}
\cellcolor{white}
& \shortstack[l]{Directed\\reachability}
& \scopevalue{2.72}{2.41, 3.01}
& \scopevalue{4.93}{4.07, 5.76}
& \scopevalue{100.00}{100.00, 100.00} \\

\rowcolor{colorgray}
\cellcolor{white}
& \shortstack[l]{Full policy\\reevaluation}
& \scopevalue{0.67}{0.51, 0.83}
& \scopevalue{2.84}{2.47, 3.22}
& \scopevalue{97.17}{96.76, 97.57} \\

\rowcolor{white}
\cellcolor{white}
\multirow{-5}{*}{%
    \rotatebox{90}{\shortstack{\textbf{Mfg.}\\\textbf{energy}}}}
& \textbf{SCOPE-DT}
& \scopevalue{0.66}{0.49, 0.82}
& \scopevalue{3.63}{3.05, 4.21}
& \scopevalue{96.35}{95.72, 96.95} \\
\hline

\rowcolor{colorgray}
\cellcolor{white}
& \shortstack[l]{Global\\suspension}
& \scopevalue{0.00}{0.00, 0.00}
& \scopevalue{100.00}{100.00, 100.00}
& \scopevalue{0.00}{0.00, 0.00} \\

\rowcolor{white}
\cellcolor{white}
& \shortstack[l]{Fixed local\\isolation}
& \scopevalue{12.39}{11.09, 13.73}
& \scopevalue{5.21}{4.63, 5.80}
& \scopevalue{95.07}{94.57, 95.59} \\

\rowcolor{colorgray}
\cellcolor{white}
& \shortstack[l]{Directed\\reachability}
& \scopevalue{3.25}{2.89, 3.61}
& \scopevalue{4.87}{4.27, 5.57}
& \scopevalue{100.00}{100.00, 100.00} \\

\rowcolor{white}
\cellcolor{white}
& \shortstack[l]{Full policy\\reevaluation}
& \scopevalue{0.97}{0.71, 1.27}
& \scopevalue{2.56}{2.26, 2.87}
& \scopevalue{97.42}{97.08, 97.76} \\

\rowcolor{colorgray}
\cellcolor{white}
\multirow{-5}{*}{%
    \rotatebox{90}{\shortstack{\textbf{Banking}\\\textbf{payments}}}}
& \textbf{SCOPE-DT}
& \scopevalue{0.94}{0.69, 1.23}
& \scopevalue{3.38}{2.85, 3.93}
& \scopevalue{96.60}{96.01, 97.14} \\
\hline

\rowcolor{white}
\cellcolor{white}
& \shortstack[l]{Global\\suspension}
& \scopevalue{0.00}{0.00, 0.00}
& \scopevalue{100.00}{100.00, 100.00}
& \scopevalue{0.00}{0.00, 0.00} \\

\rowcolor{colorgray}
\cellcolor{white}
& \shortstack[l]{Fixed local\\isolation}
& \scopevalue{12.05}{10.59, 13.50}
& \scopevalue{5.41}{4.94, 5.93}
& \scopevalue{94.83}{94.30, 95.27} \\

\rowcolor{white}
\cellcolor{white}
& \shortstack[l]{Directed\\reachability}
& \scopevalue{3.24}{2.89, 3.61}
& \scopevalue{5.00}{4.17, 5.84}
& \scopevalue{100.00}{100.00, 100.00} \\

\rowcolor{colorgray}
\cellcolor{white}
& \shortstack[l]{Full policy\\reevaluation}
& \scopevalue{0.81}{0.59, 1.05}
& \scopevalue{2.51}{2.18, 2.80}
& \scopevalue{97.48}{97.13, 97.83} \\

\rowcolor{white}
\cellcolor{white}
\multirow{-5}{*}{%
    \rotatebox{90}{\shortstack{\textbf{Insurance}\\\textbf{claims}}}}
& \textbf{SCOPE-DT}
& \scopevalue{0.77}{0.58, 0.98}
& \scopevalue{3.28}{2.82, 3.81}
& \scopevalue{96.69}{96.15, 97.21} \\
\hline
\end{tabularx}

\vspace{1mm}
\parbox{\columnwidth}{%
\scriptsize
Values are mean per-run percentages with 95\% confidence
intervals below; 96 runs per domain and method.
Safe held denotes unnecessary holds among safe proposals.
}
\vspace{-3mm}
\end{table}
{\looseness=-1
Across the three domains, SCOPE-DT and full policy reevaluation produced identical unsafe-execution and unaffected-completion rates in six of eight scenarios, as both evaluate the same dependency predicates for affected actions. SCOPE-DT completion in these scenarios ranged from 97.33\% to 98.47\%. Signer compromise and compound failure were more disruptive. In these two scenarios, SCOPE-DT also applies an additional probabilistic fail-closed rule, which lowered unaffected completion to 93.24\% and 91.97\%, compared with 96.35\% and 95.36\% for full policy reevaluation, and slightly lowered unsafe execution to 2.33\% and 1.96\%, compared with 2.43\% and 2.05\%. SCOPE-DT's additional containment therefore introduces some continuity cost and does not fully eliminate unsafe execution. Fig.~\ref{fig:scenario_results} compares outcomes across all eight scenarios.
\par}

\subsection{Dependency-Bound Permit Enforcement}
\label{sec:permit_results}
{\looseness=-1
The permit evaluation comprised 24,192 checks, split evenly between SCOPE-DT and a global-epoch invalidation baseline that revokes all permits when the global epoch changes and retains them otherwise. Table~\ref{tab:permit_results} summarizes stale-acceptance rates among invalid permits and rejection rates among valid permits. For unrelated epoch changes, SCOPE-DT rejected 0.34\% of valid permits, while the baseline rejected 100\%. In alternative-support tests, where the global epoch remains unchanged, the baseline accepted all invalid permits. In contrast, SCOPE-DT accepted 0.51\% of invalid permits and rejected 0.70\% of valid permits. For relevant dependency-epoch, policy-version, and state-version changes, SCOPE-DT recorded stale-acceptance rates of 0.45\%, 0.34\%, and 0.58\%, respectively, compared to zero for the baseline. These nonzero SCOPE-DT errors indicate residual enforcement failures, modeled as a decision flip with probability $0.055e$, where $e$ is the configured graph-error rate. Although the epoch comparison in Eq.~(3) is exact, it relies on accurate dependency records.
\par}
\begin{table}[t]
\caption{Dependency-Bound Permit Enforcement Results}
\label{tab:permit_results}
\centering
\footnotesize
\setlength{\tabcolsep}{3pt}
\renewcommand{\arraystretch}{1.0}

\setlength{\aboverulesep}{0pt}
\setlength{\belowrulesep}{0pt}
\setlength{\cmidrulesep}{0pt}

\begin{tabularx}{\columnwidth}{
@{}
>{\raggedright\arraybackslash}p{0.30\columnwidth}
r
*{4}{>{\centering\arraybackslash}X}
@{}
}
\hline
\rowcolor{colorgray}
& &
\multicolumn{2}{c}{\textbf{Global epoch}} &
\multicolumn{2}{c}{\textbf{SCOPE-DT}} \\

\cmidrule(lr){3-4}
\cmidrule(lr){5-6}

\rowcolor{colorgray}
\textbf{Test category} &
\textbf{$n$} &
\textbf{SA} &
\textbf{VR} &
\textbf{SA} &
\textbf{VR} \\
\hline

\rowcolor{white}
Alternative support &
2,517 & 100.00 & 0.00 & 0.51 & 0.70 \\

\rowcolor{colorgray}
Unrelated epoch &
2,387 & -- & 100.00 & -- & 0.34 \\

\rowcolor{white}
Relevant epoch &
2,460 & 0.00 & -- & 0.45 & -- \\

\rowcolor{colorgray}
Policy version &
2,321 & 0.00 & -- & 0.34 & -- \\

\rowcolor{white}
State version &
2,411 & 0.00 & -- & 0.58 & -- \\

\hline
\end{tabularx}

\vspace{1mm}
\parbox{\columnwidth}{%
\scriptsize
$n$: checks per method.
SA: stale acceptance among invalid permits (\%).
VR: rejection among valid permits (\%).
--: no applicable permits.
}
\vspace{-3mm}
\end{table}

\subsection{Recovery and In-Flight Workflow Effects}
\label{sec:recovery_results}

The recovery evaluation includes 23,220 method-level records, with 4,644 cases per method across five workflow phases. Restoration times are generated from method-specific recovery models over the generated obligation durations. Premature restoration refers to resuming normal authority before the recorded ground-truth readiness time. SCOPE-DT recorded a mean per-run premature-restoration rate of 0.93\%, compared with 10.64\% for fixed local isolation, 9.78\% for directed reachability, 5.22\% for full policy reevaluation, and none for global suspension.

\begin{figure}[t]
\centering
\includegraphics[width=0.85\columnwidth]{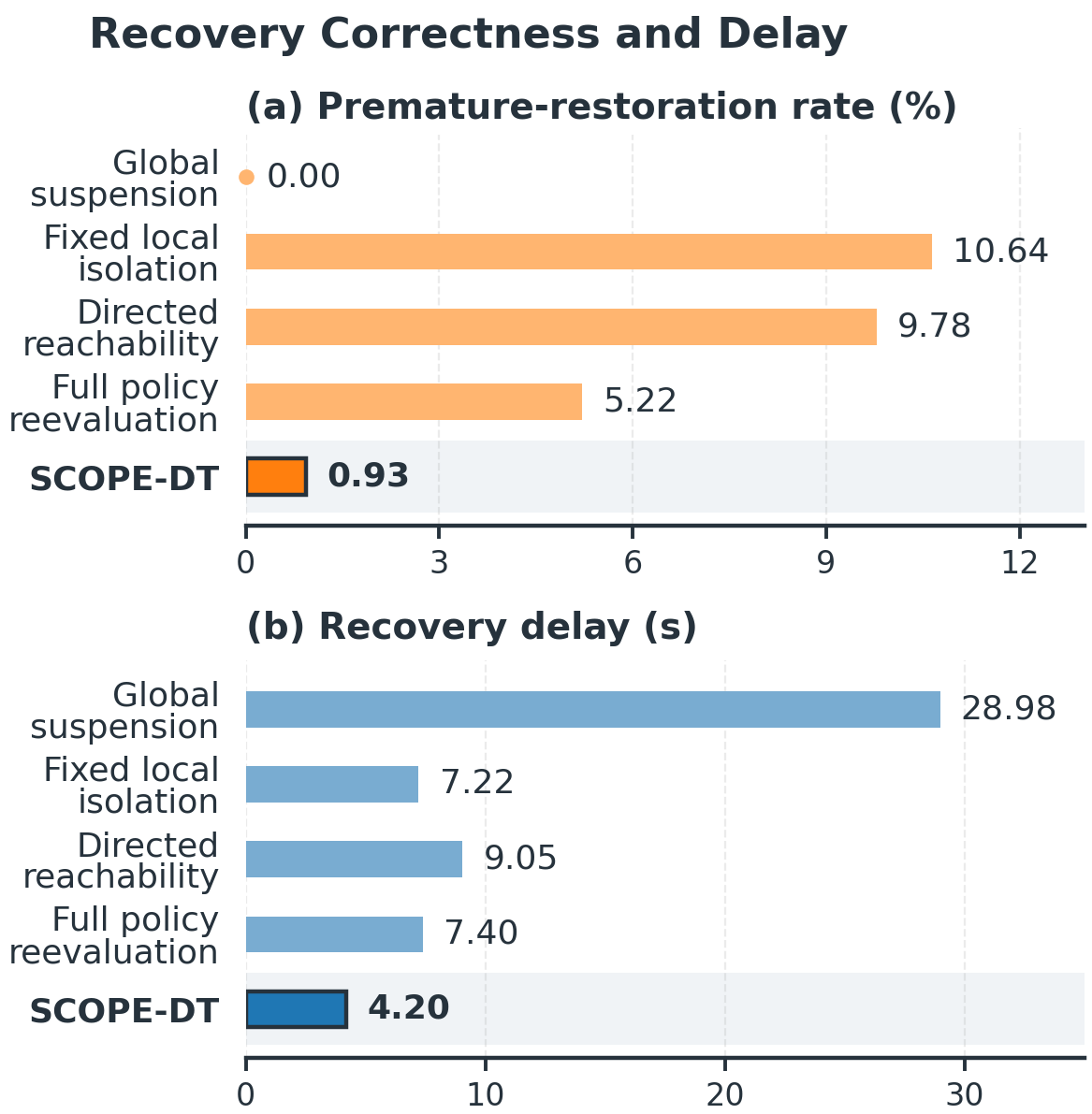}
\caption{Recovery correctness and delay across five methods. Values are means of per-run results over 288 runs per method. Premature restoration is assigned zero delay, so delay must be interpreted alongside restoration correctness.}
\label{fig:recovery_results}
\vspace{-4mm}
\end{figure}

Recovery delay measures time beyond the recorded readiness point. The mean per-run delay was 4.20~s for SCOPE-DT, compared with 7.22~s for fixed local isolation, 9.05~s for directed reachability, 7.40~s for full policy reevaluation, and 28.98~s for global suspension. Among the 284 runs with unresolved-effect cases, the mean premature-restoration rate for those cases was 1.02\% under SCOPE-DT and 5.70\% under full policy reevaluation. Fig.~\ref{fig:recovery_results} compares premature restoration and recovery delay across methods.

\subsection{Mechanism Sensitivity Analysis}
\label{sec:mechanism_sensitivity}

\begin{figure}[t]
\centering
\includegraphics[width=0.90\columnwidth]{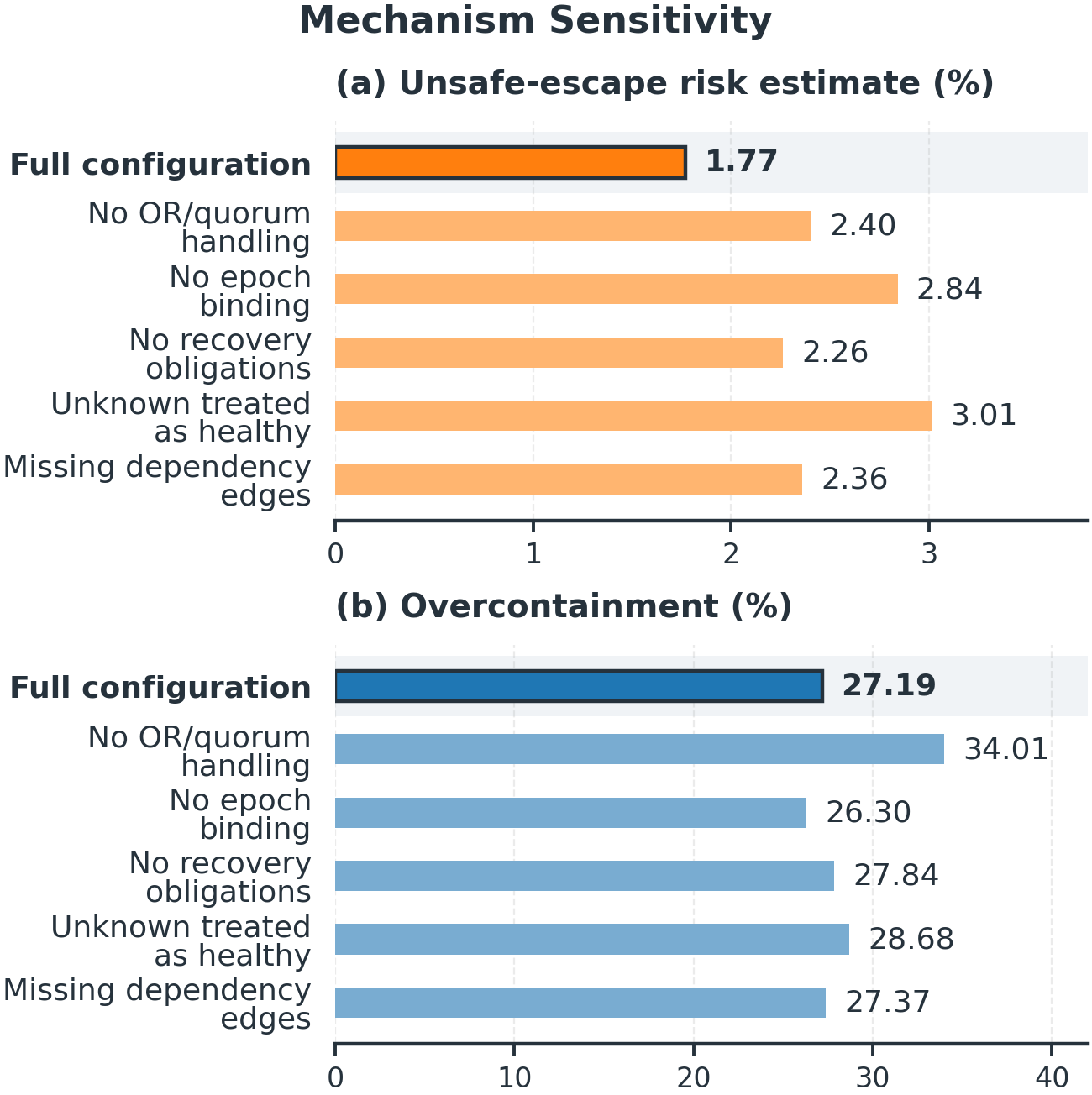}
\caption{Mechanism sensitivity across six variants. Bars show mean unsafe-escape risk estimates and overcontainment across 288 records per variant. The full configuration is highlighted.}
\label{fig:mechanism_sensitivity}
\vspace{-3mm}
\end{figure}

Fig.~\ref{fig:mechanism_sensitivity} reports 1,728 generated records across six parameterized variants. For each record, unsafe-escape risk is estimated as $100\min(0.95,Bs m_v)$, where $B\sim\mathrm{Beta}(1.2,30)$, $s$ is incident severity, and $m_v$ is the configured multiplier for variant $v$. Overcontainment is estimated separately as $100\min(0.99,Cq_v)$, where $C\sim\mathrm{Beta}(3,8)$ and $q_v=1.22$ for the variant without OR/quorum handling and $1$ otherwise. The full configuration yields a mean unsafe-escape risk of 1.77\%, compared to 2.84\% without epoch binding and 3.01\% when unknown dependencies are treated as healthy, and a mean overcontainment of 27.19\%, compared to 34.01\% without OR/quorum handling. These estimates are generated rather than obtained by disabling and rerunning mechanisms. Overcontainment is therefore distinct from the safe-held rates in Table~\ref{tab:containment_results}, which count ground-truth-safe proposals denied by each method. The analysis does not estimate premature restoration for the variant without recovery obligations, which Section~\ref{sec:recovery_results} compares across methods.

\subsection{Scalability and Computational Cost}
\label{sec:scalability}

\begin{figure*}[t]
\centering
\includegraphics[width=0.87\textwidth]{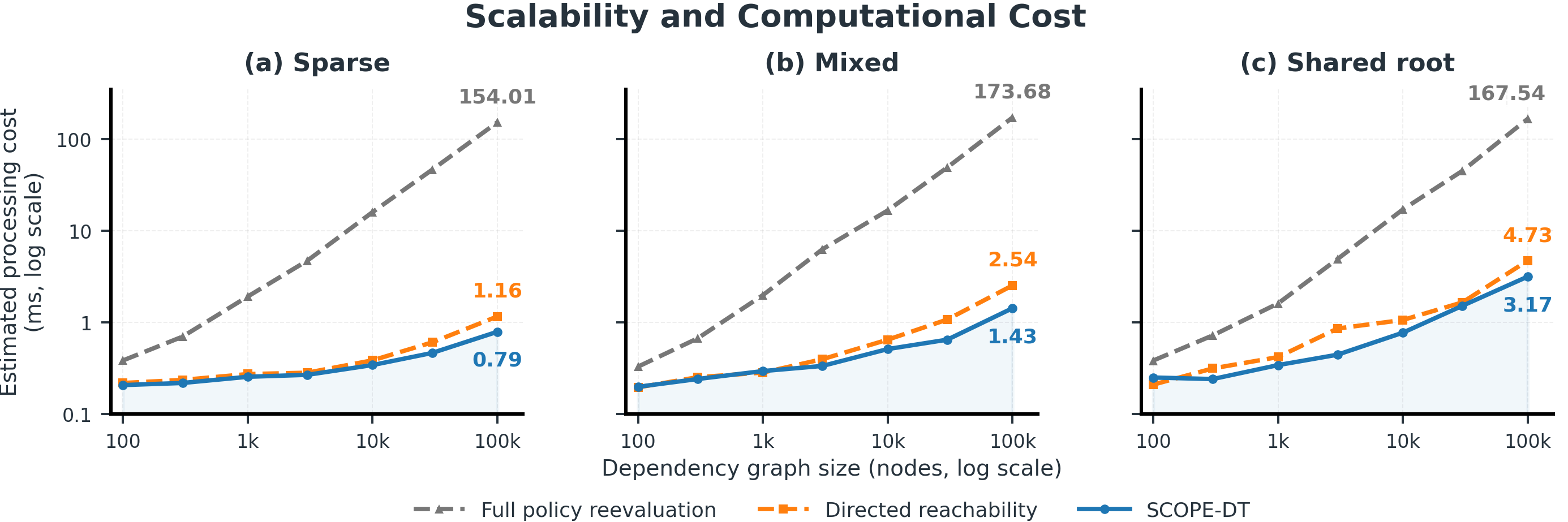}
\caption{Estimated processing cost across graph sizes and topologies. Each point is the mean of 35 repetitions using the configured cost function. Both axes use logarithmic scales; annotations show values at 100,000 nodes.}
\label{fig:scalability_results}
\vspace{-3pt}
\end{figure*}

The scaling analysis includes 2,205 records across seven graph sizes (100 to 100,000 nodes), three topologies, and three methods, with 35 repetitions per combination. Processing costs are estimated from a configured cost function that assigns each method a per-node coefficient. For graphs with 100,000 nodes, SCOPE-DT's mean estimated processing cost ranged from 0.79 to 3.17~ms across topologies, compared with 1.16--4.73~ms for directed reachability and 154.01--173.69~ms for full policy reevaluation. SCOPE-DT and directed reachability reevaluated an estimated mean of 1,310 to 5,991 nodes, depending on topology, while full policy reevaluation examined all 100,000 nodes. The difference between SCOPE-DT and directed reachability at equal node counts therefore reflects the configured coefficients rather than measured processing time. Shared-root topology produced the largest affected sets and the highest estimated SCOPE-DT cost. Fig.~\ref{fig:scalability_results} compares processing-cost estimates across graph sizes and topologies.

\section{Industrial Adoption}
\label{sec:industrial_adoption}

Industrial adoption requires integration with existing operational systems and clear ownership of dependency records and recovery policies. Standards-based digital-twin architectures support integration and interoperability~\cite{kim2022implementation}, while industrial experience highlights the need to adapt twin implementations to application-specific requirements~\cite{tao2024advancements}. For SCOPE-DT, deployment would involve mapping operational dependencies, connecting validated observations to the registry, and establishing enforcement points at effectful service interfaces. A phased rollout could start with monitoring and operator-reviewed restrictions, followed by automated containment. Domain owners would define acceptable evidence sources, compensation rules, and reconciliation criteria. Table~\ref{tab:industrial_adoption} outlines proposed adoption requirements for the three evaluated domains and for hospital operations.

\begin{table*}[t]
\caption{Industry-Specific Adoption Requirements for SCOPE-DT}
\label{tab:industrial_adoption}
\centering
\footnotesize
\setlength{\tabcolsep}{4pt}
\renewcommand{\arraystretch}{0.9}

\begin{tabular}{
@{}
>{\raggedright\arraybackslash}p{0.12\textwidth}
>{\raggedright\arraybackslash}p{\dimexpr0.26\textwidth-2\tabcolsep\relax}
>{\raggedright\arraybackslash}p{\dimexpr0.32\textwidth-2\tabcolsep\relax}
>{\raggedright\arraybackslash}p{\dimexpr0.30\textwidth-2\tabcolsep\relax}
@{}
}
\hline
\rowcolor{colorgray}
\textbf{Industry} &
\textbf{Operational dependencies} &
\textbf{Selective containment} &
\textbf{Recovery requirements} \\
\hline

\rowcolor{white}
Manufacturing and energy~\cite{shao2021usecases} &
Equipment state, sensor freshness, and shared energy-service availability &
Restrict affected equipment commands or scheduling actions while preserving independently supported operations &
Renew evidence, verify equipment and service readiness, and confirm restart conditions \\

\rowcolor{colorgray}
Banking payments~\cite{bcbs2021resilience} &
Account state, transaction authorization, and payment-service availability &
Hold payment instructions with unsupported requirements while preserving independent payment processing &
Reconcile uncertain transfers and verify account and settlement state before retry or continuation \\

\rowcolor{white}
Insurance claims~\cite{eiopa2021aigovernance} &
Policy coverage, claim documents, external assessments, and payout status &
Hold unsupported claim decisions or payouts while allowing independently supported processing &
Renew supporting evidence, resolve inconsistent assessments, and reconcile uncertain payout outcomes \\

\rowcolor{colorgray}
Hospital operations~\cite{han2023smarthospital} &
Bed occupancy and readiness, staffing capacity, transfer orders, and acceptance records &
Hold affected bed reservations or transfer steps while preserving independently supported coordination &
Renew occupancy and staffing evidence, verify receiving acceptance, and reconcile uncertain transfer status \\

\hline
\end{tabular}

\vspace{1mm}
\parbox{\textwidth}{
\scriptsize
References provide domain context. The containment and recovery mappings are proposed applications of SCOPE-DT. Hospital operations was not included in the evaluation.
}
\vspace{-5pt}
\end{table*}
\section{Discussion}
\label{sec:discussion}

\textbf{Safety and Operational Continuity.}
The results suggest that dependency-scoped containment reduces unsupported execution more effectively than fixed local isolation and directed reachability, while preserving most unaffected operations. However, full policy reevaluation achieved slightly higher unaffected completion. Signer compromise and compound failures produced the largest continuity penalties, underscoring the significance of shared trust dependencies and incident interactions. Operational decisions should therefore consider unsafe execution, unnecessary restriction, and restoration correctness together. SCOPE-DT showed lower premature restoration, but did not fully eliminate it.

\textbf{Limitations.}
SCOPE-DT relies on accurate dependency records, reliable evidence, and consistent enforcement at effectful-step boundaries. Missing edges, misclassified requirements, or compromised sources may weaken containment effectiveness. The evaluation is limited to three domains and eight generated incident scenarios. Processing costs are derived from a configured cost function rather than measurements of implementation overhead.

\textbf{Future Work.}
Planned work includes measuring end-to-end enforcement latency and assessing recovery during overlapping incidents. Causal tracing across components could inform dependency maintenance~\cite{mace2015pivot}, and linearizability checking could assess the ordering of concurrent incident updates and step admission~\cite{burckhardt2010lineup}. Deployment studies will also examine operator intervention, integration with existing authorization and workflow services, and extension to other domains such as hospital operations.
\section{Conclusion}
\label{sec:conclusion}

SCOPE-DT connects action-level execution authority to operational dependencies, dependency epochs, and incident-specific recovery obligations. The evaluation shows reduced unsupported execution compared with fixed local isolation and directed reachability while preserving most unaffected operations. Under the evaluated recovery models, SCOPE-DT showed lower premature restoration than the other selective methods, although residual failures remain. Full policy reevaluation retains slightly higher continuity, highlighting a trade-off between selective restriction and operational availability. The framework provides a basis for integrating containment and recovery into agentic digital twins.


\balance
\bibliographystyle{IEEEtran}
\bibliography{references}

\end{document}